\documentclass[conference]{IEEEtran}
\IEEEoverridecommandlockouts
\usepackage{cite}
\usepackage{amsmath,amssymb,amsfonts}
\usepackage{algorithmic}
\usepackage{graphicx}
\usepackage{textcomp}
\usepackage{xcolor}
\usepackage{enumitem}
\usepackage{caption}
\usepackage{subcaption}
\usepackage{array}
\usepackage{url}
\def\BibTeX{{\rm B\kern-.05em{\sc i\kern-.025em b}\kern-.08em
    T\kern-.1667em\lower.7ex\hbox{E}\kern-.125emX}}
\begin{document}

\graphicspath{{figures/}}

\title{Dissecting the Performance of Chained-BFT}
\author{
\IEEEauthorblockN{Fangyu Gai\IEEEauthorrefmark{1}, Ali Farahbakhsh\IEEEauthorrefmark{1}, Jianyu Niu\IEEEauthorrefmark{1}, Chen Feng\IEEEauthorrefmark{1}, Ivan Beschastnikh\IEEEauthorrefmark{2}, Hao Duan\IEEEauthorrefmark{3}}
\IEEEauthorblockA{University of British Columbia (\IEEEauthorrefmark{1}Okanagan Campus, \IEEEauthorrefmark{2}Vancouver Campus)}
\IEEEauthorblockA{\IEEEauthorrefmark{3}Hangzhou Qulian Technology Co., Ltd.}
}


\maketitle
\thispagestyle{plain}
\pagestyle{plain}

\begin{abstract}
Permissioned blockchains employ Byzantine fault-tolerant (BFT) state machine replication (SMR) to reach agreement on an ever-growing, linearly ordered log of transactions.
A new paradigm, combined with decades of research in BFT SMR and blockchain (namely chained-BFT, or cBFT), has emerged for directly constructing blockchain protocols.
Chained-BFT protocols have a unifying propose-vote scheme instead of multiple different voting phases with a set of voting and commit rules to guarantee safety and liveness.
However, distinct voting and commit rules impose varying impacts on performance under different workloads, network conditions, and Byzantine attacks. Therefore, a fair comparison of the proposed protocols poses a challenge that has not yet been addressed by existing work.

We fill this gap by studying a family of cBFT protocols with a two-pronged systematic approach.
First, we present an evaluation framework, Bamboo, for quick prototyping of cBFT protocols and that includes helpful benchmarking facilities.
To validate Bamboo, we introduce an analytic model using queuing theory which also offers a back-of-the-envelope guide for dissecting these protocols.
We build multiple cBFT protocols using Bamboo and we are the first to fairly compare three representatives (i.e., HotStuff, two-chain HotStuff, and Streamlet).
We evaluated these protocols under various parameters and scenarios, including two Byzantine attacks that have not been widely discussed in the literature.
Our findings reveal interesting trade-offs (e.g., responsiveness vs. forking-resilience) between different cBFT protocols and their design choices, which provide developers and researchers with insights into the design and implementation of this protocol family.
\end{abstract}


\section{Introduction}
Classic Byzantine fault-tolerant (BFT) state machine replication (SMR) protocols like PBFT~\cite{PBFT} rely on a stable leader to drive the protocol until a view change occurs.
This limits their scalability and deployability in a context that involves thousands of nodes and requires highly distributed trust. 
Recent work has been exploring an alternative, called \textit{chained-BFT} or cBFT, combining decades of research in BFT and state-of-the-art blockchain work, which is considered to be the next generation BFT for blockchains.

\begin{figure}[t]
    \begin{center}
      \includegraphics[width=\linewidth]{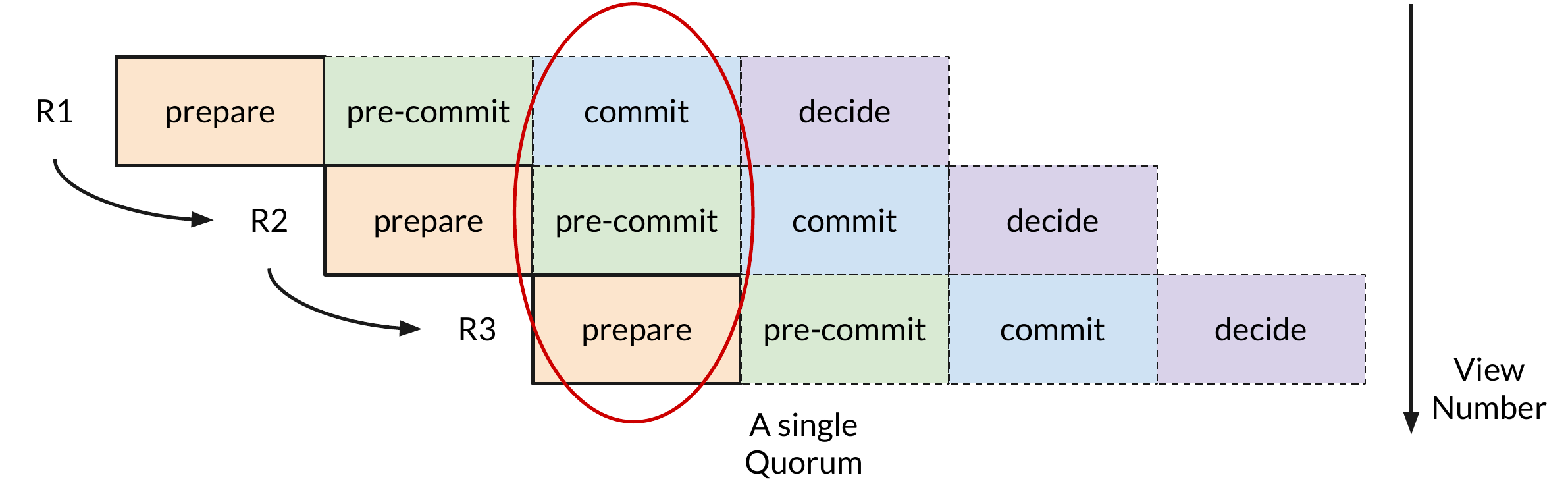}
      \caption{Run-time of chained-HotStuff~\cite{hotstuff}. Replicas $R_1,R_2,$ and $R_3$ are rotating to propose blocks. A QC accumulated at a phase for a single block can serve in different phases of the preceding blocks.}
      \label{fig:pipeline}
    \end{center}
\end{figure}

The run-time of a cBFT protocol is divided into views.
Each view has a designated leader, which is elected via some leader election protocol.
Unlike classical BFT protocols, such as PBFT~\cite{PBFT}, in which each proposed block has to go through three different phases of voting, cBFT protocols use the chained structure that enables a single-phase \textit{Propose-Vote} scheme.
Since blocks are cryptographically linked together, a vote cast on a block is also cast on the preceding blocks of the same branch.
This allows for streamlining of decisions and fewer messages types, and also eases the burden of reasoning about the correctness of the protocols.
See an illustration of chained HotStuff in Figure~\ref{fig:pipeline}.
Votes for the same block from distinct replicas are accumulated as a \textit{Quorum Certificate} (short for QC) if a threshold is reached (i.e., over two thirds of all the nodes) and the QC is also recorded on the blockchain along with the relevant block for bookkeeping.

Liveness and safety properties are guaranteed by voting and commit rules, both of which characterize a cBFT protocol.
Informally, a \emph{voting rule} requires that an incoming block is voted on only if it extends a valid branch, while a \emph{commit rule} requires that a chain of blocks is committed by removing a certain number of blocks at the tail of the blockchain.
However, the choice of rules changes performance under different conditions: network environment (latency and bandwidth), workload, deployment parameters, and failures.

The idea of connecting modern blockchains with classic BFT originates implicitly in Casper-FFG~\cite{casper}, followed by a series of recent proposals~\cite{dfinity,casper,pala,hotstuff,streamlet,fasthotstuff,sft}.
Many companies have been using these protocols.
For example, Diem from Facebook (previously known as Libra), Flow from Dapper Labs, and Hyperchain from Qulian have deployed cBFT protocols as their core infrastructure for payment, gaming, and supply chain systems~\cite{Novi,Dapper,Qulian}.
Understanding the performance of these protocols is important for the overall distributed system.
However, these protocols differ in their voting and commit rules and design choices.
To our knowledge, there has been no study that empirically provides a fair and comprehensive comparison of these protocols.

To dissect and explain the performance of chained-BFT variants, we propose a novel framework called \textit{Bamboo}.
Bamboo provides well-defined interfaces so that developers can quickly prototype their own cBFT protocols by defining voting/commit rules.
We used Bamboo to build multiple cBFT protocols, including HotStuff~\cite{hotstuff}, two-chain HotStuff, Streamlet~\cite{streamlet}, Fast-HotStuff~\cite{fasthotstuff}, and LBFT~\cite{lbft}.
We then comprehensively evaluated the performance of three of these protocols when subjected to different consensus and network parameters, including two Byzantine attacks that could lead to blockchain forking.
Since our protocol implementations share most components, Bamboo provided a fair comparison, especially on the impact of different voting/commit rule choices.

A challenge we face when engineering Bamboo is finding the right modularity boundary to express a variety of BFT protocol designs.
In Bamboo, we divide a protocol into three modules: \emph{data}, \emph{pacemaker}, and \emph{safety}.
The data module maintains a block forest so that blocks can be added and pruned efficiently.
The pacemaker module ensures liveness: it allows slow nodes to catch up to the latest view so that the protocol can always make progress.
The safety module is for making decisions when new messages arrive.
Particularly, the safety module defines the \textbf{Proposing} rule (i.e., how to make a block proposal), \textbf{Voting} rule (i.e., whether to vote for an incoming block), \textbf{State Updating} rule (i.e., how the state should be updated), and the \textbf{Commit} rule (i.e., whether to commit a set of blocks).
The data and pacemaker modules are shared by many protocols, leaving the safety module to be specified by developers.

To cross-validate our Bamboo BFT implementations, we devised an analytical model using queuing theory~\cite{queue}.
This model estimates the latency and the throughput of a cBFT protocol implementation.
We show that the results from experiments executed with Bamboo align with our mathematical model, which further indicates that our model can be used to dissect the performance of a cBFT protocol and provide a back-of-the-envelope performance estimate.
In summary, we make the following contributions:
\begin{itemize}
    \item We present the first prototyping and evaluation framework, called Bamboo, for building, understanding, and comparing chained-BFT protocols.
    We have released the framework for public use\footnote{\url{https://github.com/gitferry/bamboo}}.
    \item We introduce an analytical model using queuing theory, which verifies our Bamboo-based implementations and provides a back-of-the-envelope performance estimation for cBFT protocols.
    \item We conduct a comprehensive evaluation of three representative cBFT protocols: HotStuff, two-chain HotStuff, and Streamlet.
    Our empirical results capture the performance differences among these protocols under various scenarios including two Byzantine cases that have not been widely discussed in the literature.
    The results serve as a baseline for further development of permissioned blockchain protocols.
\end{itemize}

\section{Chained-BFT Protocols}\label{sec:background}
In this section, we present the family of chained-BFT (cBFT) SMR protocols. We first provide an overview of cBFT to outline the basic machinery and then introduce protocol details. Our descriptions of the protocols are informal.
Please see a more formal description of the protocols in a concurrent work~\cite{sft}.


\subsection{Overview}
At a high level, cBFT protocols share a unifying propose-vote paradigm in which they assign transactions coming from the clients a unique order in the global ledger. This ledger, also known as a blockchain, is a sequence of blocks cryptographically linked together by hashes. Each block in a blockchain contains a hash of its parent block along with a batch of transactions and other metadata.  

\begin{figure}[t]
    \begin{center}
      \includegraphics[width=0.9\linewidth]{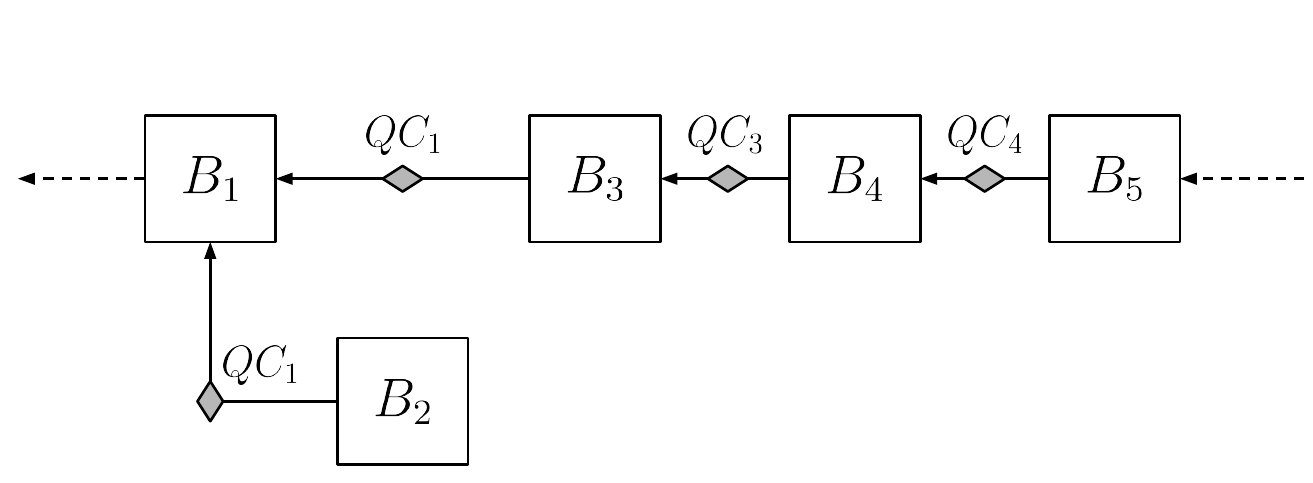}
      \caption{The tree structure of cBFT-based blockchains.
      The subscription of a block $B$ is referred to the view of the block being proposed while that of a $QC$ is referred to the block that the \textit{QC} certifies.
      Note that forks can happen due to the network delay and malicious behavior.}
      \label{fig:propose-vote}
    \end{center}
\end{figure}

Similar to classic BFT protocols, cBFT protocols are driven by leader nodes and operate in a view-by-view manner.
Each participant takes actions on receipt of messages according to four protocol-specific rules: \textbf{Proposing}, \textbf{Voting}, \textbf{State Updating}, and \textbf{Commit}.
Each view has a designated leader chosen at random, which proposes a block according to the \textbf{Proposing} rule and distributes the block through the blockchain network.
On receiving a block, replicas take actions according to the \textbf{Voting} rule and update their local state according to the \textbf{State Updating} rule.
For each view, replicas should certify the validity of the proposed block by forming a \textit{Quorum Certificate} (or QC) for the block.
A block with a valid QC is considered certified.
The basic structure of a blockchain is depicted in Figure~\ref{fig:propose-vote}.
Forks can happen because of conflicting blocks, which is a scenario in which two blocks do not extend each other.
Conflicting blocks might arise because of network delays or proposers deliberately ignoring the tail of the blockchain.

Replicas finalize a block whenever the block satisfies the \textbf{Commit} rule based on their local state.
Once a block is finalized, the entire prefix of the chain is also finalized. Rules dictate that all finalized blocks remain in a single chain.
Finalized blocks can be removed from memory to persistent storage for garbage collection.

\subsection{HotStuff}\label{sec:hotstuffintro}
\begin{figure}[t]
    \begin{center}
      \includegraphics[width=0.9\linewidth]{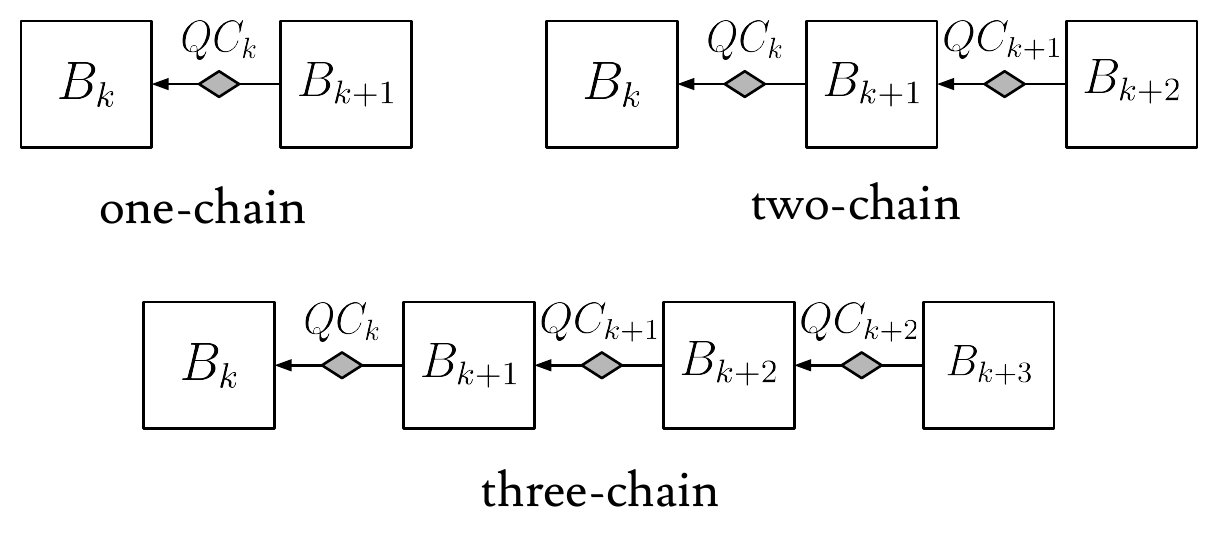}
      \caption{Illustration of HotStuff's one-chain, two-chain, and three-chain.}
      \label{fig:commit-rule}
    \end{center}
\end{figure}

The HotStuff protocol~\cite{hotstuff} is considered as the first BFT SMR protocol that achieves linear view change and optimistic responsiveness\footnote{Optimistic responsiveness means that a correct leader is guaranteed to make progress at network speed without waiting for some apriori upper bound on network delay.}.
These two properties combine to make frequent leader rotation practical, which is crucial for fairness and liveness.
Before describing the four rules behind HotStuff, we first introduce some definitions.
HotStuff defines a \textbf{one-chain} as a block certified by a QC, a \textbf{two-chain} as a one-chain that has a direct descendent one-chain, and a \textbf{three-chain} as a two-chain whose tail block is certified, illustrated in Figure~\ref{fig:commit-rule}.

\textbf{State Updating} rule.
We first define some important state variables.
The locked block ($lBlock$) is the head of the highest two-chain.
The last voted view ($lvView$) is the highest view in which the voting rule is met.
The highest QC ($hQC$) is the highest QC collected from received messages.
The variables $lBlock$ and $hQC$ are updated when a new QC is received, while $lvView$ is updated right after a vote is sent.

\textbf{Proposing} rule.
When a replica becomes the leader for a certain view, it builds a block based on $hQC$ and broadcasts the proposal.

\textbf{Voting} rule.
When a replica sees an incoming block $b^*$, it checks to see (1) whether $b^*.view$ is larger than $lvView$, and (2) whether it extends $lBlock$ ($lBlock$ is $b^*$'s grandparent block) or its parent block has a higher view than that of $lBlock$.
The vote is sent to the leader of the next view when the two conditions are met.
We ignore semantic checks for brevity.

\textbf{Commit} rule.
HotStuff follows a three-chain commit rule to make decisions on blocks to be committed.
The head of the three-chain along with all the preceding blocks in the same branch are committed as long as a three-chain emerges.
As illustrated in Figure~\ref{fig:propose-vote}, when $b_{v4}$ becomes certified (on receiving $QC_{v4}$), $b_{v1}$ is not committed since $b_{v3}$ is not its directed descendent one-chain.
As soon as $b_{v5}$ is certified (on receiving $QC_{v5}$), $b_{v3}$, $b_{v1}$, and all their preceding blocks become committed.

HotStuff decouples liveness guarantee into a module called \textit{Pacemaker} which is used to synchronize sufficiently many honest nodes into the same view if they happen to be out of sync.
We leave details of this process to Section~\ref{sec:pacemaker}.

\subsection{Two-chain HotStuff}
Two-chain HotStuff (2CHS) is a two-phase variant of HotStuff which could commit a block after two rounds of voting, similar to Tendermint~\cite{Tendermint} and Casper~\cite{casper}.
The state variables are the same as in HotStuff except that in the \textbf{State Updating} rule, $lBlock$ is the head of the one-chain.
The \textbf{Voting} rule and \textbf{Proposing} rule remain the same with HotStuff.
The \textbf{Commit} rule of 2CHS requires a two-chain.
Although the 2CHS saves one round of voting which leads to lower latency as compared to HotStuff, it is not responsive.
Since the lock is on one-chain, the proposal has to be built on top of the highest one-chain to make progress.
To achieve progress, leaders must wait for the maximal network delay to collect messages from all the honest replicas after a view change.

\subsection{Streamlet}\label{sec:streamlet}
Streamlet is proposed for pedagogy due to its simplicity.
The four rules of Streamlet are based on a similar principle of the longest chain rule from Bitcoin.
The state of Streamlet is a \textit{notarized} chain which is a chain of certified blocks.
The \textbf{State Updating} rule is to maintain such a \textit{notarized} chain.

\textbf{Proposing} rule.
The leader proposes a block built on top of the longest \textit{notarized} chain.

\textbf{Voting} rule.
A replica will vote for the first proposal, only if the proposed block is built on top of the longest \textit{notarized} chain it has seen.
Note that the vote is broadcast.

\textbf{Commit} rule. Whenever three blocks proposed in three consecutive views get certified, the first two blocks out of the three along with the ancestor blocks are committed.

The original Streamlet protocol requires a synchronized clock and each view has a duration of $2\Delta$, which is considered to be the maximum network delay.
To fairly compare protocols, we modify Streamlet in Bamboo by replacing the synchronized clock with the same Pacemaker component described in Section~\ref{sec:pacemaker}.
The validity of this replacement has been previously discussed in~\cite{Dahlia}.
Note that in Streamlet, all the messages are echoed and the votes are broadcast, incurring $O(n^3)$ communication complexity.
Additionally, although Streamlet also has a three-chain commit rule, it does not enjoy optimistic responsiveness as it still requires timeouts to ensure liveness.

\section{Bamboo Design}\label{sec:bamboo}
\begin{figure}[t]
    \begin{center}
      \includegraphics[width=\linewidth]{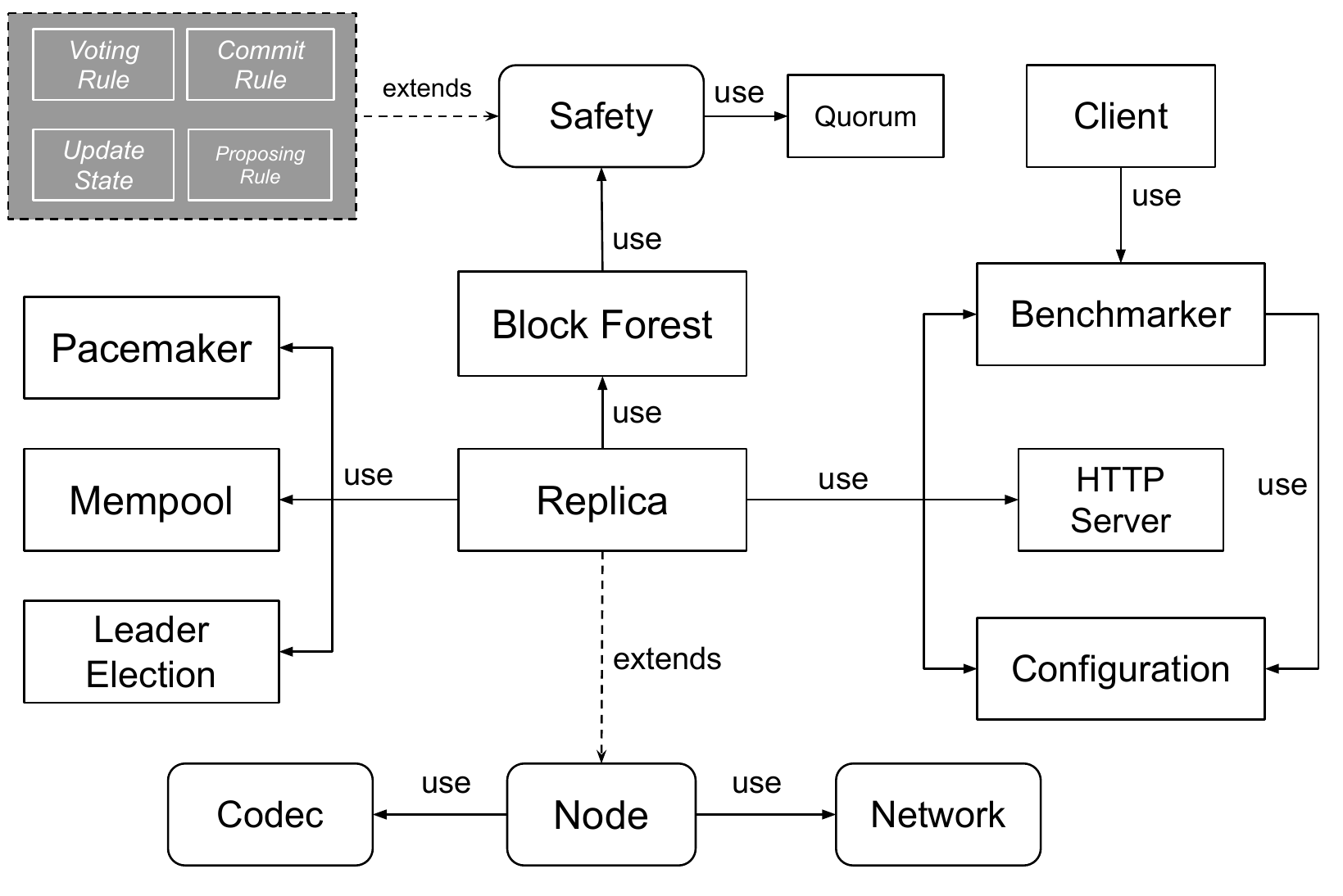}
      \caption{Modular design of the Bamboo framework.}
      \label{fig:modules}
    \end{center}
\end{figure}

In this section, we present the Bamboo's design. 
We start with the observation that cBFT protocols share common components, such as the blockchain data structure, message handlers, and network infrastructures.
Bamboo offers implementations for these shared parts.
Each component resides in its own loosely-coupled module and exposes a well-defined API.
This design enables modules to be extended and replaced to accommodate different design choices.
Figure~\ref{fig:modules} overviews Bamboo's architecture.
Developers can easily prototype a cBFT-based blockchain protocol by filling in the \textbf{voting}, \textbf{commit}, \textbf{State Updating}, and \textbf{Proposing} rules (shaded blocks) that extend the \textit{Safety} API.
Using the benchmark facilities, users can conduct comprehensive apple-to-apple comparisons across different cBFT protocols and their design choices.
Next, we describe the major components of Bamboo.

\subsection{Block Forest}
We use the block forest module to keep track of blocks.
Block forest contains multiple block trees, which is a potentially disconnected planar graph.
Each vertex in the graph is assigned a height, which increases monotonically.
A vertex can only have one parent with a strictly smaller height but it can have multiple children, all with strictly larger height.
A block forest provides the ability to prune all vertices up to a specific height.
A tree whose root is below the pruning threshold might decompose into multiple disconnected sub-trees as a result of pruning.
The Block Forest component guarantees that there is always a main branch, or \textit{main chain}, which contains all the committed blocks cryptographically linked in the proposed order.
As a result, a consistency check can be easily conducted by checking the hash of the block at the same height across nodes.

\subsection{Pacemaker}\label{sec:pacemaker}
The pacemaker module is in charge of advancing views.
It encapsulates the view synchronization logic that ensures liveness of cBFT protocols.
Specifically, view synchronization aims to keep a sufficient number of honest replicas in the same view for a sufficiently long period of time so that collaboration will continue and progress can always be made.
This module is first defined in HotStuff~\cite{hotstuff} without detailed specification.
LibraBFT~\cite{libra} provides an implementation of this module which we adopt in our framework.
The general idea is essentially the same as a Byzantine fault detector~\cite{detector}.
In this module, whenever a replica times out at its current view, say $v$, it broadcasts a timeout message $\left \langle \textsc{Timeout, $v$}\right \rangle$ and advances to view $v+1$ as soon as a quorum ($2f+1$) of timeout messages of $v$, called \textit{TimeoutCertificate} (or TC), is received.
In addition, the TC will be sent to the leader of view $v+1$ which will drive the protocol forward if it is honest.
Note that the Pacemaker module causes quadratic message complexity even though HotStuff has linear message complexity in the happy path.
Other proposals such as Cogsworth~\cite{cogsworth} present optimized solutions that have reduced message complexity and constant latency, which will be considered in future work.

\subsection{Safety Module}
The safety module defines all the interfaces needed to implement the consensus core.
It consists of the voting rule, commit rule, state updating rule, and the proposing rule.
Developers implement these rules according to protocol specifications.
The voting and commit rules are key to guaranteeing safety and liveness, as described in Section~\ref{sec:background}.
On receipt of incoming messages, the state variables are updated to ensure that the voting and commit rules can be eventually met.
The proposing rule defines how a block is proposed in the hope that it can make progress. We implement the two Byzantine strategies (see Section~\ref{sec:byzantine}) by modifying the proposing rule.

\begin{table}[h!]
  \begin{center}
    \caption{Major configuration parameters.}
    \label{tab:parameters}
    \begin{tabular}{|m{5em}|m{3em}|m{17em}|}
    \hline
      \textbf{Parameter} & \textbf{Default Value} & \textbf{Description}\\
    \hline
    address & nil & List of peers in which key is the ID and value is the internal IP\\
    \hline
    master & 0 & ID of the static leader; rotating if it is set to 0\\
    \hline
    strategy & silence & Byzantine strategy\\
    \hline
    byzNo & 0 & Number of Byzantine nodes\\
    \hline
    bsize & 400 & Number of transactions included in a block\\
    \hline
    memsize & 1000 & Number of transactions held in the memory pool\\
    \hline
    psize & 0 & Payload size of a transaction (bytes)\\
    \hline
    delay & 0 & Additional delay of messages sent\\
    \hline
    timeout & 100 & Waiting time for entering the next view (ms)\\
    \hline
    runtime & 30 & The period clients run for\\
    \hline
    concurrency & 10 & Number of concurrent clients\\
    \hline
    \end{tabular}
  \end{center}
\end{table}

\subsection{Benchmark Facilities}
The configuration, client library, and benchmarker components are the core of Bamboo's capability for evaluating cBFT protocols.
Bamboo is a configurable framework for both nodes and clients.
The configuration component provides configuration parameters about the environment; these are listed in Table~\ref{tab:parameters}.
A configuration is fixed for each run and managed via a JSON file distributed to every node.

The Bamboo client library uses a RESTful API to interact with server nodes, which enables the system to run any benchmark, such as YCSB~\cite{ycsb} and Hyperledger Caliper~\cite{caliper}.
The benchmarker generates tunable workloads and measures the performance of protocols in terms of throughput, latency, and the two other metrics, which we discuss in the following section.
Bamboo also supports simulating network fluctuation and partitioning by tuning the ``delay'' parameter, which can also be tuned during run-time by sending a ``slow'' command to a certain node.
Since Bamboo currently focuses on protocol-level performance, we adopt an in-memory key-value data store for simplicity.
We will consider a VM-based execution layer in our future work.

\subsection{Other components}

\textbf{Mempool.} This component serves as a memory pool which is a bidirectional queue in which new transactions are inserted from the back while old transactions (from forked blocks) are inserted from the front.
Each node maintains a local memory pool to avoid duplication check.

\textbf{Network.} Bamboo reuses network module from Paxi~\cite{Paxi} which is implemented as a simple message-passing model.
The transport layer supports TCP and Go channel for communication, which allows both large-scale deployment and single-machine simulation.

\textbf{Quorum.} A quorum system is a key component for ensuring consistency in any distributed fault-tolerant systems.
This component supports two simple interfaces to collects votes (via the interface \textit{voted()}) and generate QCs (via \textit{certified()}).

\section{Byzantine Attacks and Metrics}

Since it is impossible to enumerate all the Byzantine attacks~\cite{byzantine}, we consider two simple Byzantine attacks that are specific to cBFT protocols and target their performance.
We assume that metrics are measured after the GST (Global Stabilization Time), which means we do not consider network-level attacks from an outsider, such as eclipse attack~\cite{eclipes} and Denial-of-Service attack, which may cause a network partition (i.e., asynchronous network) and block progress.

\subsection{Attack Strategies}\label{sec:byzantine}

We introduce two attack strategies that are challenging to detect as the attackers are not violating the protocol from an outsider's view, but could damage performance.
Superficially, both attacks may cause forks.
Developers can easily implement these attack strategies in less than 50 LoC of Go code in Bamboo by modifying the Proposing Rule.
We emphasize that both Byzantine strategies are not optimal, and finding an optimal strategy is beyond the scope of this paper.

\subsubsection{Forking Attack}

\begin{figure}[h]
    \begin{center}
      \includegraphics[width=0.9\linewidth]{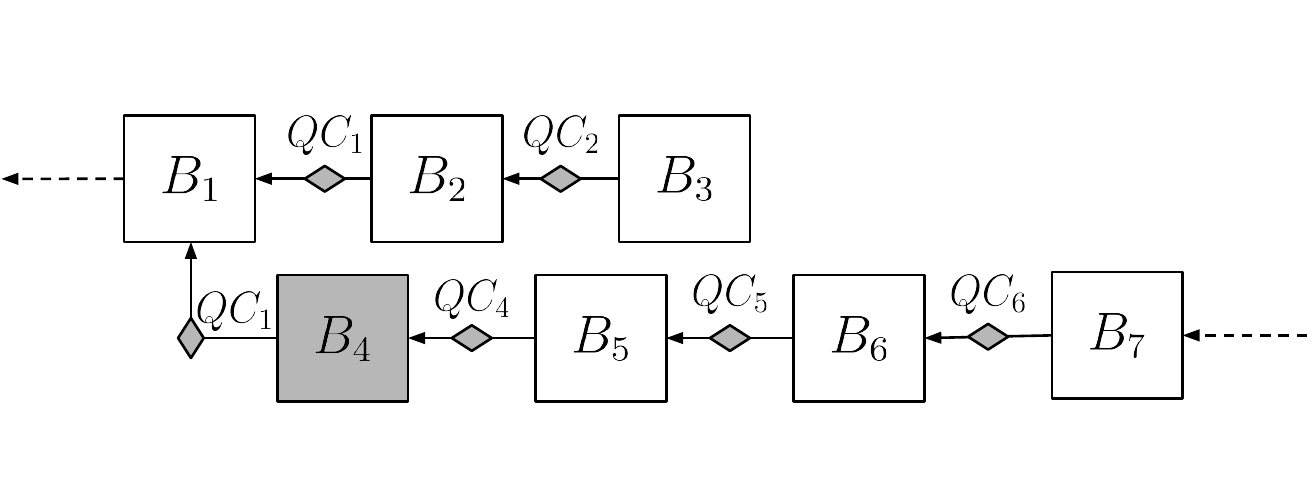}
      \caption{An example of forking attack in HotStuff.}
      \label{fig:forking-attack}
    \end{center}
\end{figure}

The forking attack aims to overwrite\footnote{overwritten block will be simply ignored and garbage-collected} previous blocks that have not been committed.
Attackers perform forking attack by proposing conflicting blocks.
Take an example from HotStuff, shown in Figure~\ref{fig:forking-attack}. The leader for view 4 has launched a forking attack by building $B_{4}$ on top of $B_{1}$ instead of $B_{3}$.
Although $B_{4}$ branches out from the main chain, this is a valid proposal from the view of other replicas according to the voting rule.
Therefore, replicas will cast votes on $B_{4}$ and honest leaders afterward will continue to build blocks following this branch.
Eventually, $B_{4}$ becomes committed and $B_{2}$ and $B_{3}$ are overwritten.
Similarly, the two-chain HotStuff is also subject to the forking attack, except that the attacker can only create a fork up to $B_{2}$ because of the constraint of the voting rule.
On the contrary, Streamlet is resilient to this attack because votes are broadcast and replicas only vote for blocks that are built following the longest chain.

\subsubsection{Silence Attack}

\begin{figure}[h]
    \begin{center}
      \includegraphics[width=0.9\linewidth]{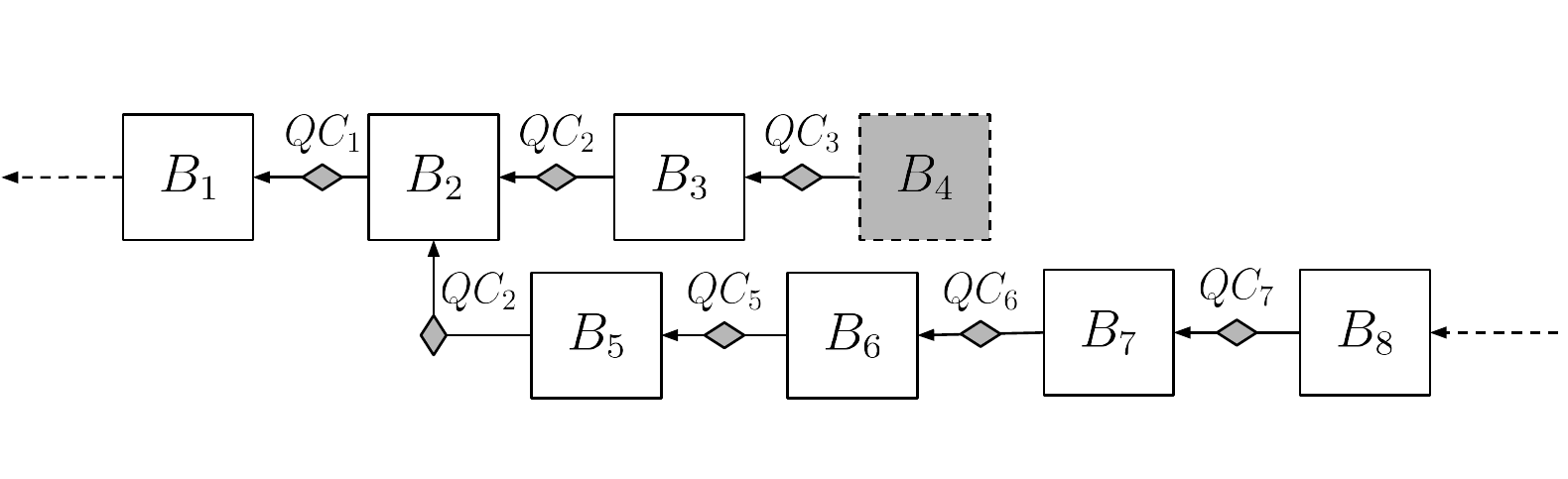}
      \caption{An example of silence attack in HotStuff.}
      \label{fig:silence-attack}
    \end{center}
\end{figure}

The purpose of the silence attack is to break the commit rule, thus extending the time before a block becomes committed.
One way to launch this attack is for the attacker to simply remain silent when it is selected as the leader, until the end of the view.
An example of this attack is shown in Figure~\ref{fig:silence-attack}, in which $B_{4}$ is withheld in view 4 and the (honest) leader for view 5 has to build $B_{5}$ following $B_{2}$ because of the loss of $QC_{3}$.
Recall that a block is committed if and only if three consecutive blocks extend it.
Thus $B_{1}$, which could have been committed in view 4, becomes committed in view 8 ($QC_{7}$ is attached to $B_{8}$).
At the same time, $B_{3}$ is overwritten.
Note that the silence attack will trigger timeouts, which could lead to long delay, depending on timeout settings.

\subsection{Metrics}\label{sec:metrics}
Bamboo measures performance in terms of latency, throughput, and two other micro-metrics, \emph{chain growth rate} and \emph{block intervals}.
Throughput is the number of transactions that are processed per second, or $Tx/s$, while latency is measured at the client end, and refers to the time between the client sending the transaction and receiving a confirmation.
Since it is important to show the performance of a protocol in terms of tail latency under stress, these two metrics are measured by increasing the benchmark throughput (by increasing the concurrency level of the clients) until the system is saturated.

To have a deeper understanding of the impact of the two Byzantine attacks mentioned earlier, we introduce two micro metrics, \textit{chain growth rate} and \textit{block interval}.
\begin{enumerate}
    \item \textbf{Chain growth rate ($CGR$):} From a replica's view, the chain growth rate is defined as the rate of committed blocks appended onto the blockchain over the long run.
    Let $C(v)$ denote the number of committed blocks during $v$ views.
    Thus, we have:
    \begin{equation}
        CGR=\lim_{v\to\infty} \frac{C(v)}{v}
    \end{equation}
    \item \textbf{Block interval ($BI$):} The block interval $BI$ measures the average number of views a block takes from being added to the blockchain until being committed over the long run.
    Let $L_i$ denote the number of views that the $i$-th block takes from the view $b_i$ is produced to the view that the block is committed during $v$ views.
    Thus, we have:
    \begin{equation}
        BI=\lim_{v\to\infty} \frac{\sum_{i=1}^{C(v)} L_i}{C(v)}
    \end{equation}
\end{enumerate}

\section{Performance Model}
In this section, we introduce a mathematical model that allows us to estimate the delay performance of cBFT protocols.
This model is similar in spirit to the queuing model proposed in~\cite{Paxi} except that we have different leader-election and commit rules (which complicate the analysis).
This model will be used later to validate the Bamboo's implementation.

\subsection{Assumptions}

We use the same set of assumptions as used in \cite{Paxi} about the machines, network, and transactions. 

\subsubsection{Machines}
We assume that all the machines (being used to host replicas or clients) have the same network bandwidth and identical CPUs. 
For simplicity, we focus on the case that each machine has a \emph{single} CPU and NIC.
We also assume that each replica has a unique local memory pool.


\subsubsection{Network}
We consider a system of $N$ nodes and assume that the Round-Trip Time (RTT) in the network between any two nodes follows a normal distribution with mean $\mu$ and standard deviation $\sigma$. (In practice, $\mu$ and $\sigma$ can be determined via measurement.) This assumption is justified for a LAN setting in \cite{Paxi}. 

\subsubsection{Transactions}
For simplicity, we assume that transactions are of equal size so that each block contains the same (maximum) number of transactions denoted by $n$. Also, we assume that the transactions arrive into the system according to a Poisson process with rate $\lambda$.


\subsection{Building Blocks}

Before we conduct our delay analysis, we introduce some building blocks. There are two types of delays in the system: machine-related delay and network-related delay.

\subsubsection{Machine-related delay}

Similar to \cite{Paxi}, we treat each machine as a single queue consisting of CPU and NIC. The CPU delay, denoted by $t_{CPU}$, captures the delay of operations such as signing and verifying signatures. Since the machines have identical CPUs, $t_{CPU}$ is a constant parameter.

The NIC delay, denoted by $t_{NIC}$, is the total amount of time a block spends in the NIC of the sender and the receiver. 
Let us denote the bandwidth of a machine as $b$ and the size of a block as $m$.
Then, we have $t_{NIC} = \frac{2m}{b}$. The factor of $2$ comes from the fact that a block first goes through the sender's NIC and then the receiver's NIC.

\subsubsection{Network-related delay}\label{sec:network-delays}
The network-related delay can be divided into two parts. First, we have the round trip delay between the client and its associated replica from which it will receive the response to its request. This delay is a normal random variable with mean $\mu$ as explained before. Second, we have another delay that captures the amount of time needed \emph{in the network} for a leader to collect a quorum of votes from the replicas. 
Since a quorum consists of 
$\frac{2N}{3}$ votes, this delay is equal to the $(\frac{2N}{3}-1)$-th order statistics of $N-1$ i.i.d. normal random variables with mean $\mu$ and standard deviation $\sigma$. Here, the term of $-1$ is necessary because the leader already has its own vote, and we just need to analyze the rest of the votes needed for a quorum. The expected value of this delay, denoted by $t_{Q}$, can be easily obtained through numerical methods (e.g., \cite{numMethods}). Alternatively, it can be obtained through Monte Carlo simulation as suggested in \cite{Paxi}.


\subsection{Delay Analysis}\label{delay-analysis}
We are now ready to present our delay analysis for cBFT protocols.
We will start with the HotStuff protocol~\cite{hotstuff}.
Our analysis will focus on the lifetime of a block in the system,
which can be used to approximate the lifetime of its transactions\footnote{This approximation is valid since a collection of transactions waiting to be included into a block can be viewed as a ``virtual'' block waiting to be proposed.}. We will not consider the effect of Byzantine attacks in our delay analysis, as it has already been discussed in our previous work \cite{HotStuffGit}. 
Instead, we will focus on the happy-path (assuming a synchronous network and no Byzantine behavior) performance of cBFT protocols (especially HotStuff), because such analysis is important for us to validate our implementation and is missing in \cite{HotStuffGit}.



\begin{figure}[h]
     \begin{center}
       \includegraphics[width=\linewidth]{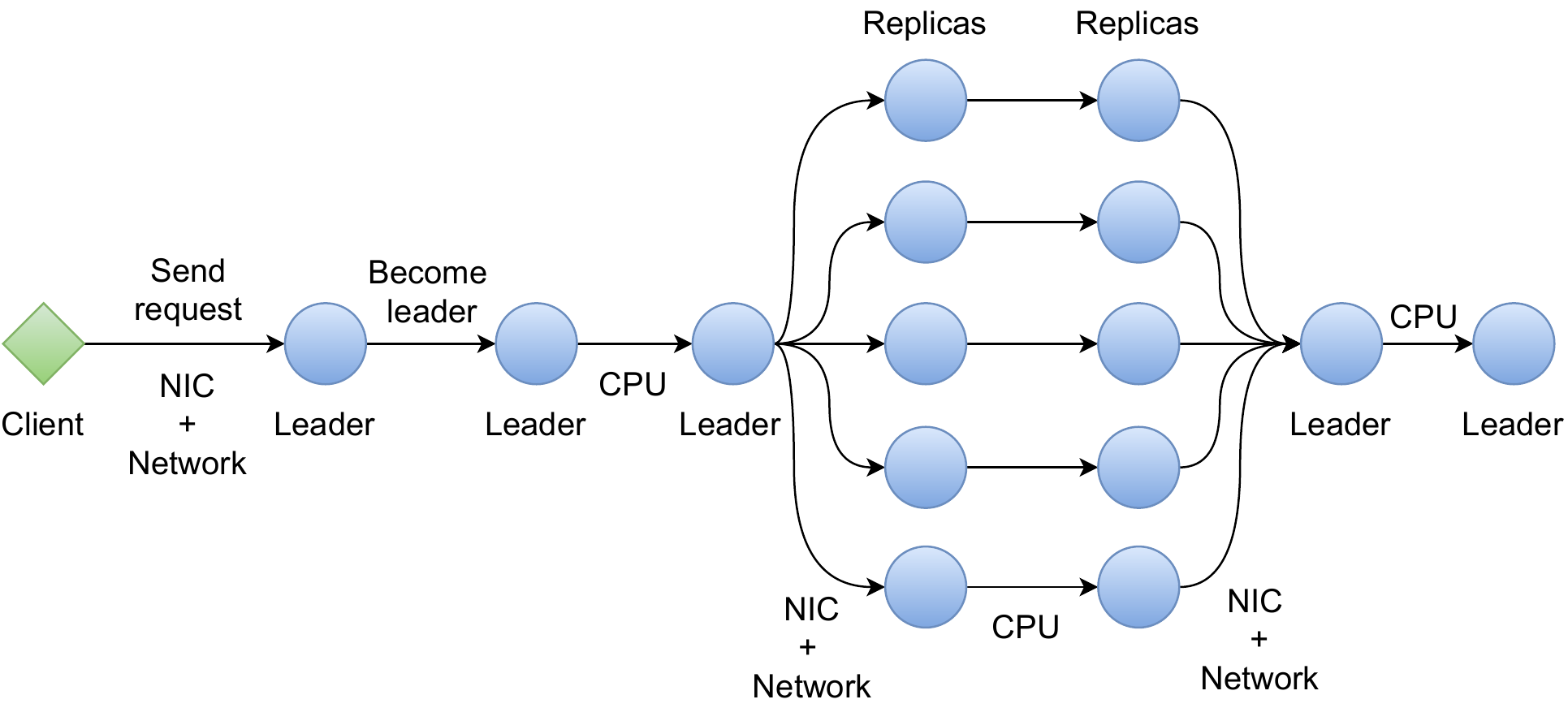}
       \caption{Transaction lifetime in HotStuff, excluding $t_{commit}$}
       \label{fig:HotStuffModel}
     \end{center}
\end{figure}

Generally, the latency of a transaction can be divided into the following parts as illustrated in Figure~\ref{fig:HotStuffModel}:
\begin{equation}
latency = t_{L}  + t_{s} + t_{Commit} + w_{Q},
\label{Eq:R}
\end{equation}
where $t_{L}$ is the average RTT (Round Trip Time) between the client (who issues this transaction) and its associated replica,
$t_{s}$ is the average service time required to serve the block containing this transaction, $t_{Commit}$ is the average amount of time it takes for the block to get committed once it has been certified, and $w_{Q}$ is the average waiting time for the block to get processed. We will now derive these terms one by one.

\subsubsection{Derivation of $t_{L}$}
As explained in Section~\ref{sec:network-delays}, $t_{L}$ is equal to $\mu$, a pre-defined parameter from measurement.

\subsubsection{Derivation of $t_{s}$}\label{sec:t_s}
First, the block containing this transaction has to go through the CPU of the leader, and then through the NICs of the leader and the NIC of each replica the leader will send the proposal.
This takes $t_{CPU} + t_{NIC}$ units of time.
Second, the block has to wait for a quorum of votes to be gathered by the next leader. Such delay is explained in
Section~\ref{sec:network-delays} with expected value $t_Q$.
In addition, we have the CPU delay of the replicas, and the NIC delay between the next leader and the replicas, given by $t_{CPU} + t_{NIC}$.
Finally, once a quorum of votes is received by the next leader, it has to process this quorum, which takes $t_{CPU}$ units of time. Therefore, we have the following expression for $t_s$:
\begin{equation}\label{EQ:t_s}
\begin{aligned}
    t_{s} &= t_{CPU} + t_{NIC} + t_{Q} + t_{CPU} + t_{NIC} + t_{CPU} \\
    &= 3t_{CPU} + 2t_{NIC} + t_{Q}.
\end{aligned}
\end{equation}

\subsubsection{Derivation of $t_{commit}$}
Recall that a block in HotStuff is committed once a three-chain is established. Therefore, in a happy-path condition, a block would have to wait for two other blocks to get certified (see Section~\ref{sec:hotstuffintro} for details) so that the three of them would form a three-chain. Thus, we have $t_{commit} = 2t_{s}$.

\subsubsection{Derivation of $w_{Q}$}

Recall that transactions arrive into the system according to a Poisson process with rate $\lambda$.
Also recall that we assume that each transaction goes to a replica chosen uniformly at random.
Thus, transactions arrive at each replica according to a Poisson process with rate $\frac{\lambda}{N}$. 
Approximately, we can think that blocks arrive at each replica according to a Poisson process with rate $\frac{\lambda}{nN}$, where $n$ is the maximum number of transactions contained in a block.

As explained before, the average service time at each replica is given by $t_s$.
Recall that a replica becomes a leader every $N$ rounds (views) on average and only a leader can propose a block. Hence, the effective average service time (of proposing and processing a block) at each replica is $N t_s$. Therefore, the queuing process at a replica can be approximated by an M/D/1 model with the average waiting time given by
\begin{equation}
w_{Q} = \frac{\rho}{2u(1-\rho)},    
\label{Eq:WQ}
\end{equation}
where $u = \frac{1}{N t_s}$ is the effective service rate, $\rho = \frac{\gamma}{u}$, and $\gamma = \frac{\lambda}{nN}$ is the block arrival rate.

\subsection{Analysis of Streamlet and Two-Chain HotStuff}

It is straightforward to adapt the above analysis to Streamlet and Two-Chain HotStuff.
Equation~\eqref{Eq:R} stays the same for these two protocols since it captures the essence of a block's lifetime. 
The delay $t_{L}$ is a network-related parameter and is thus the same across different protocols.
The waiting time $w_{Q}$ also has the same formula, because we 
can still use an M/D/1 model to approximate the queuing process.
Therefore, we shall only focus on $t_{s}$ and $t_{commit}$.

\subsubsection{Two-Chain HotStuff}
Since the views of HotStuff and Two-chain HotStuff are identical, $t_{s}$ in Two-chain HotStuff can be calculated by using Equation~\eqref{EQ:t_s}.
The only difference between these two protocols is the fact that Two-chain HotStuff requires a two-chain to commit a block, whereas HotStuff requires a three-chain.
Therefore, we have $t_{commit} = t_{s}$ in Two-chain HotStuff (instead of $t_{commit}=2t_{s}$ in HotStuff).

\subsubsection{Streamlet}
The commit rule of Streamlet says that a block has to wait for one more block in the happy path so that it can be committed.
Thus, we have $t_{commit} = t_{s}$ in Streamlet. Next, let us turn to $t_{s}$.
Recall that the replicas broadcast their votes in Streamlet and they also echo every vote they have received.
So, in order to calculate $t_{s}$, we need to find out how long it takes for the next leader to receive a quorum of votes, including the processing and NIC delays.
This calculation can be done with different levels of complexity based on the underlying assumptions.
In the simplest case, we can assume that the next leader receives the quorum of votes from the voters themselves.
In this case, the delay analysis for $t_{s}$ will be the same as that in Section~\ref{sec:t_s}.
However, there is a chance that the next leader receives an echo of a vote even earlier than the vote itself. Intuitively, such an event makes the actual $t_{s}$ smaller than what is calculated using Equation~\eqref{EQ:t_s}. In other words, our previous analysis
in Section~\ref{sec:t_s} is still valid as an upper bound. This upper bound turns out to be fairly tight according to the experimental results we will present.


\subsection{Discussion}

The protocols we analyze have certain differences that are not necessarily apparent in the model.
Streamlet has a cubic communication complexity, and this increases the network and NIC delays (e.g., because of congestion).
These differences  are captured by the measurements of system parameters such as $\mu$ and $t_{cpu}$.
\par
Additionally, our analysis can be generalized for studying other design choices.
Two examples are given below.
\begin{itemize}
  \item The clients may choose to broadcast transactions instead of sending them to a single replica.
    \item The leader election procedure can be done differently (e.g., based on hash functions).
\end{itemize}
This allows us to explore the effect of these design choices.



\section{Experimental Results}\label{sec:evaluation}
In this section, we first present a comparison between our model and Bamboo's implementation.
Next, we compare the basic performance of the three protocols with different block sizes, payload sizes, and network delays, followed by a scalability test.
Then, we present protocols' tolerance to the two Byzantine attacks discussed in Section~\ref{sec:byzantine}.
We finish the section by showing how Bamboo can be used to evaluate protocol responsiveness. 

We use Bamboo to conduct experiments to explore the performance of three representative chained-BFT protocols: HotStuff (HS), two-chain HotStuff (2CHS), and Streamlet (SL) because they have clear trade-offs in different scenarios.
For instance, HS trades off latency (one more round of voting than 2CHS) for responsiveness while Streamlet compromises message complexity in exchange for simplicity and resilience against forking.
We aim to empirically show these trade-offs using Bamboo.

Bamboo is implemented in Golang with around 4,600 LoC and each protocol is around 300 LoC including two Byzantine strategies.
Bamboo uses secp256k1 for all digital signatures in both votes and quorum certificates.
We carry out our experiments on Tencent Cloud S5.2XLARGE16 instances with 8 vCPUs and 16 GB of RAM.
Each replica is initiated on a single VM.
We use 2 VMs as clients to send requests to randomly chosen replicas.
The VMs are located in the same data center with inter-VM latency below 1ms.
Clients issue requests to a randomly selected replica, which replicates the requests when it becomes the leader.
Requests are processed using in-memory storage.
We do not throttle the bandwidth in any run.

\begin{figure}[t]
    \begin{center}
      \includegraphics[width=1.07\columnwidth]{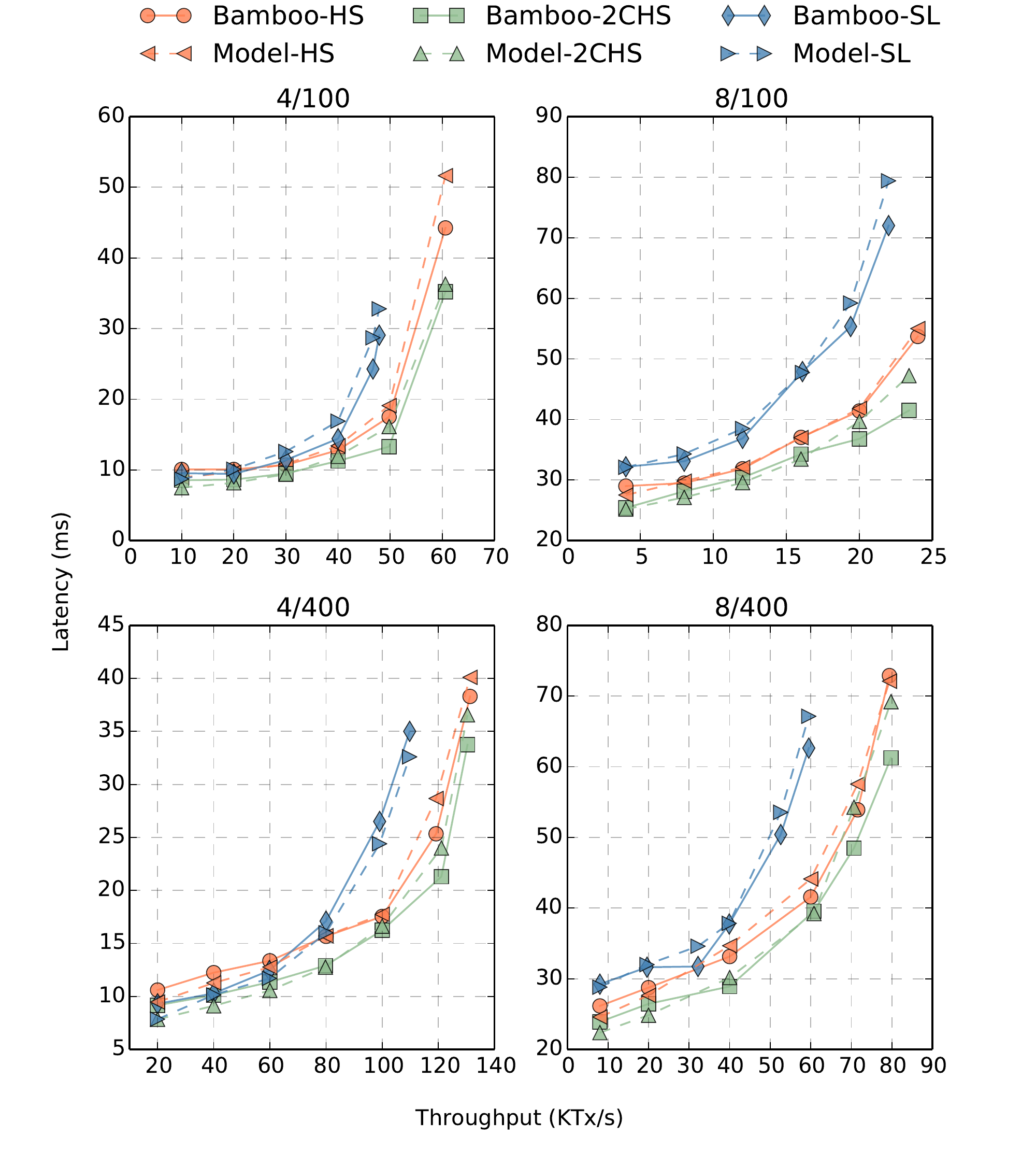}
      \caption{Comparison of the model and implementation for HotStuff, two-chain HotStuff (2CHS), and Streamlet.}
      \label{fig:model}
    \end{center}
\end{figure}

\begin{table}[h!]
  \begin{center}
    \caption{Comparison of transaction arrival rate and transaction throughput in HotStuff, for block size of 400 and 4 replicas.}
    \label{tab:arrivalVSthroughput}
    \begin{tabular}{|c||c|}
    \hline
      \textbf{Arrival rate (Tx/s)} & \textbf{Throughput (Tx/s)}\\
    \hline
    19,992 & 20,115\\
    \hline
    39,976 & 39,895\\
    \hline
    59,956 & 59,934\\
    \hline
    79,880 & 79,774\\
    \hline
    99,772 & 100,159\\
    \hline
    119,236 & 119,159\\
    \hline
    131,232 & 131,275\\
    \hline
    \end{tabular}
  \end{center}
\end{table}

\subsection{Model vs. Implementation}
To validate our Bamboo-based protocol implementations, we compare the results from our model with results from the protocol implementations. 
Our experiments use different configurations for the number of nodes (network size) and the block size. 
We evaluate the following network size/block size pairs: 4/100, 8/100, 4/400, 8/400.
The other settings are set according to the values listed in Table~\ref{tab:parameters}. In each run of an experiment the clients' concurrency level is increased until the network is saturated.

Figure~\ref{fig:model} shows four plots of throughput (thousands of transactions per second) versus latency (milliseconds), one plot per network size/block size configuration.
Each plot shows a line for each of the three Bamboo-based protocol implementations and another line for corresponding models for each protocol. By comparing the model and Bamboo lines for a protocol we can see how well the model and our implementation match.

The four plots in Figure~\ref{fig:model} demonstrate that our analytical model successfully validates our Bamboo implementations.
Furthermore, it indicates that our model can be used to perform back-of-the-envelope performance forecasting of the protocols with given workloads.
It is important to note that our model estimates latency based on the transaction arrival rate.
However, our experimental measurements indicate that the transaction throughput (throughput of transactions as observed on the blockchain) is roughly equal to the transaction arrival rate (Table~\ref{tab:arrivalVSthroughput}).
This is because in most of the workloads the delay is dominated by the queue waiting time, and is (on average) roughly the same for each block, in each workload.
Additionally, in low workloads the transactions go through a somewhat idle network, causing them to face similar $t_s$ (recall that $t_s$ is a random variable).

\subsection{Happy-Path Performance}\label{sec:happy-path}
We measure happy-path performance in terms of throughput and latency.
In the following experiments, unless specified, we use 4 replicas.

\begin{figure}[t]
    \begin{center}
      \includegraphics[width=0.9\linewidth]{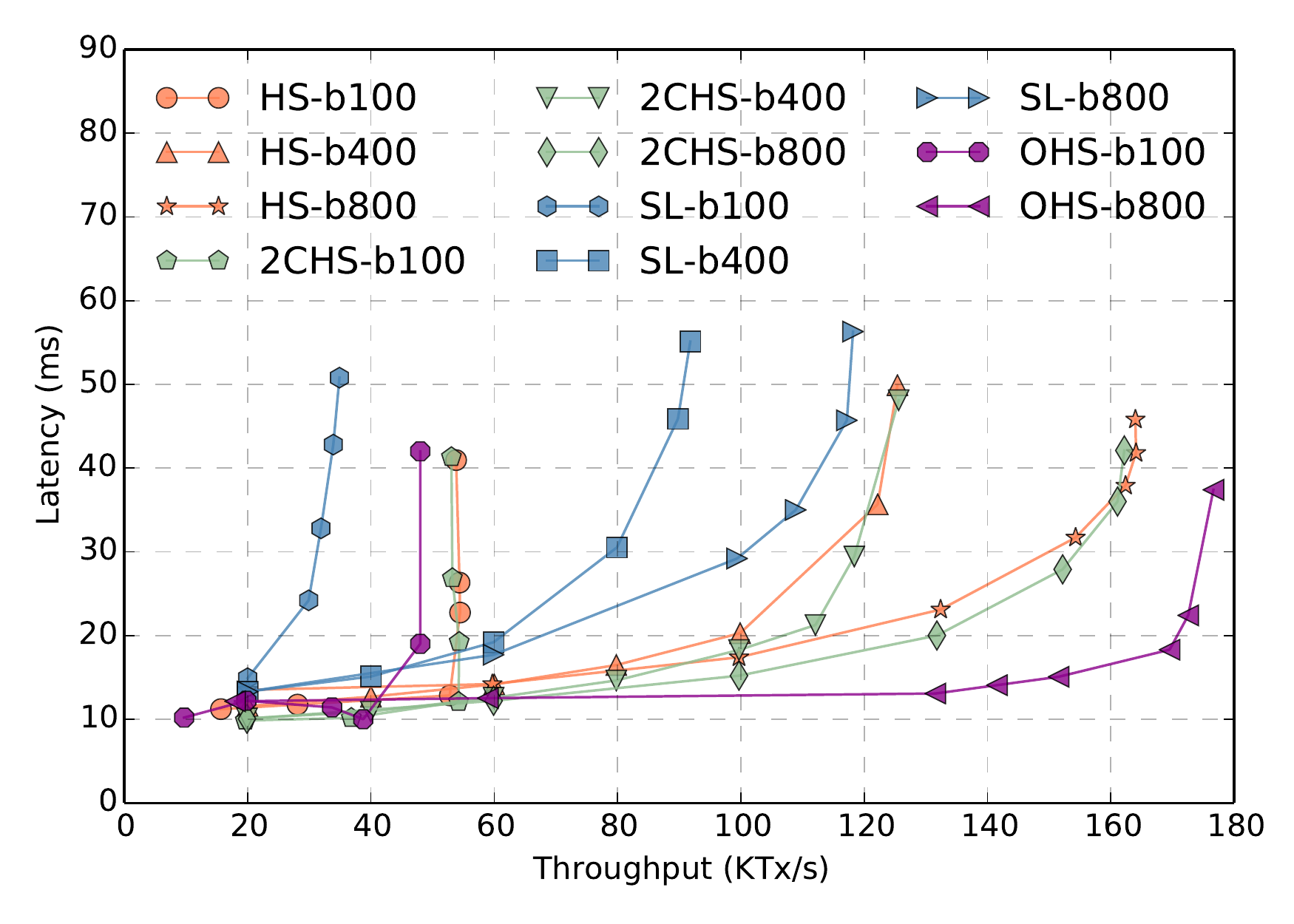}
      \caption{Throughput vs. latency with block sizes of 100, 400, and 800.}
      \label{fig:happy-path}
    \end{center}
\end{figure}

\textbf{Block sizes.}
In our first experiment, we vary the block size and compare our implementation by using the original HotStuff (OHS) implementation as the baseline.
The original HotStuff library\footnote{https://github.com/hot-stuff/libhotstuff.} is open-source and written in C\texttt{++}, allowing us to easily reproduce their basic results showed in the paper~\cite{hotstuff}.
This benchmark uses requests with zero payload and block sizes of 100, 400, and 800, denoted by ``b100'', ``b400'', and ``b800'', respectively.
We only use block sizes of 100 and 800 for OHS as we did not obtain meaningful results from the block size of 400.
Bamboo uses a simple batching strategy in which the proposer batches all the transactions in the memory pool if the amount is less than the target block size.
We gradually increase the level of client concurrency and request rate until the systems are saturated.

Figure~\ref{fig:happy-path} shows the results of the above experiment in a throughput versus latency plot, listing different protocols with varying block sizes.
We first observe that all the protocols have a typical L-shaped latency/throughput performance curve.
The Bamboo-HotStuff implementation has similar performance to the original one.
The slight gap is likely due to the original HotStuff's use of TCP instead of HTTP to accept requests from clients, different batching strategies, and language distinction.
Streamlet has lower-level throughput in all block sizes due to its broadcasting of votes and echoing of messages.
The performance gain by increasing the block size from 100 to 400 is remarkable.
However, this gain is reduced above 400 requests per block.
This is partly because the latency incurred by batching becomes higher than the cost of the crypto operations.
We use a block size of 400 in the rest of the experiments.

\begin{figure}[h]
         \includegraphics[width=0.9\linewidth]{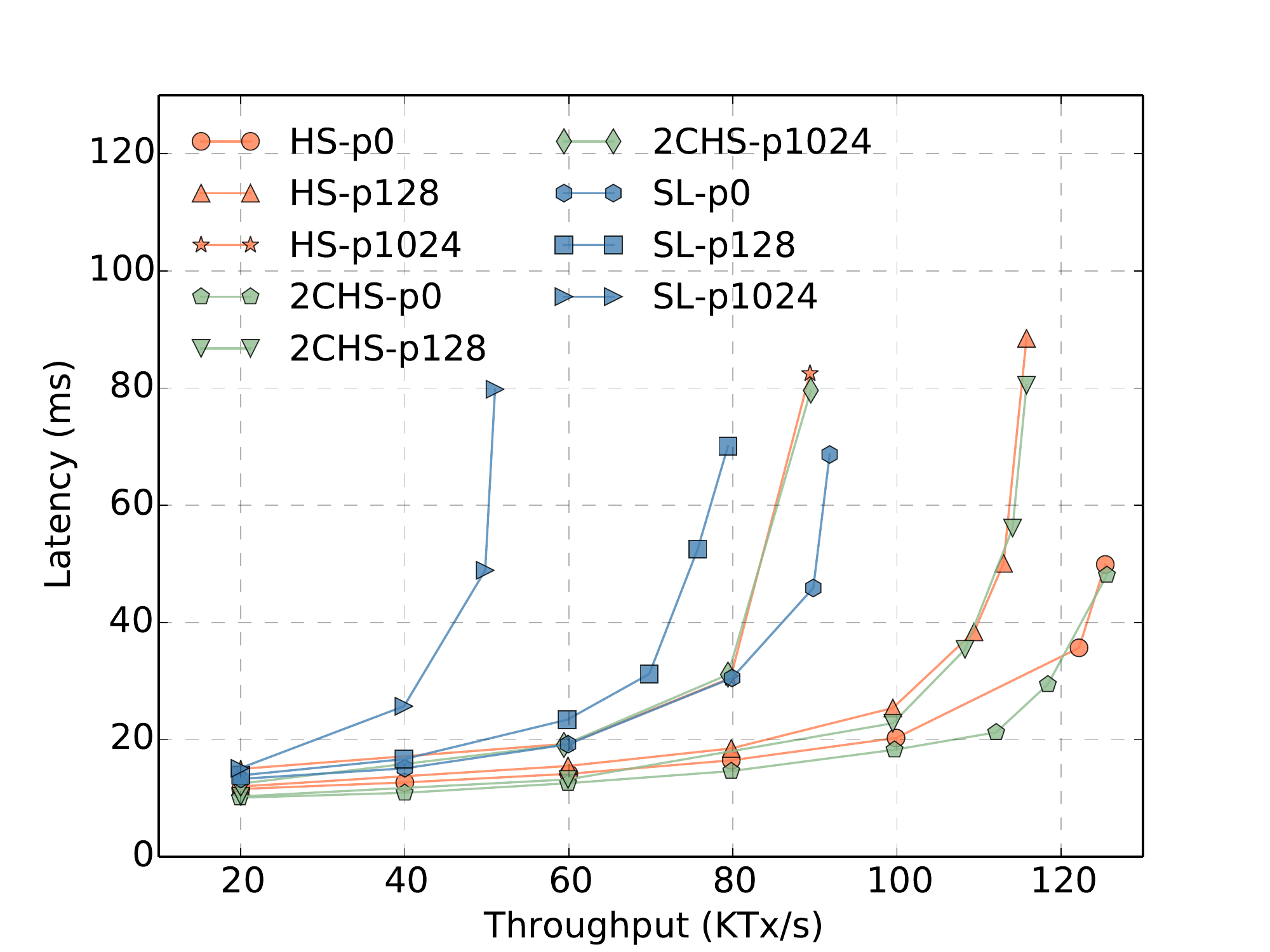}
         \caption{Throughput vs. latency with transaction payload sizes of 0, 128, and 1024 bytes.}
         \label{fig:payload-size}
\end{figure}

\textbf{Transaction payload sizes.}
Figure~\ref{fig:payload-size} illustrates three transaction payload sizes of 0, 128, 1024 (in bytes), denoted by ``p0'', ``p128'', and ``p1024'', respectively.
At all payload sizes, we can see a similar pattern for the three protocols. Streamlet has the worst performance and is more sensitive to the payload size due to its message echoing design.
The results also show that the latency difference between HotStuff and its two-chain variant becomes less obvious with a larger payload.
This is because the transmission delay of a block dominants the latency.
We set the payload size to 128 bytes in the rest of the experiments.

\begin{figure}[h]
         \includegraphics[width=0.9\linewidth]{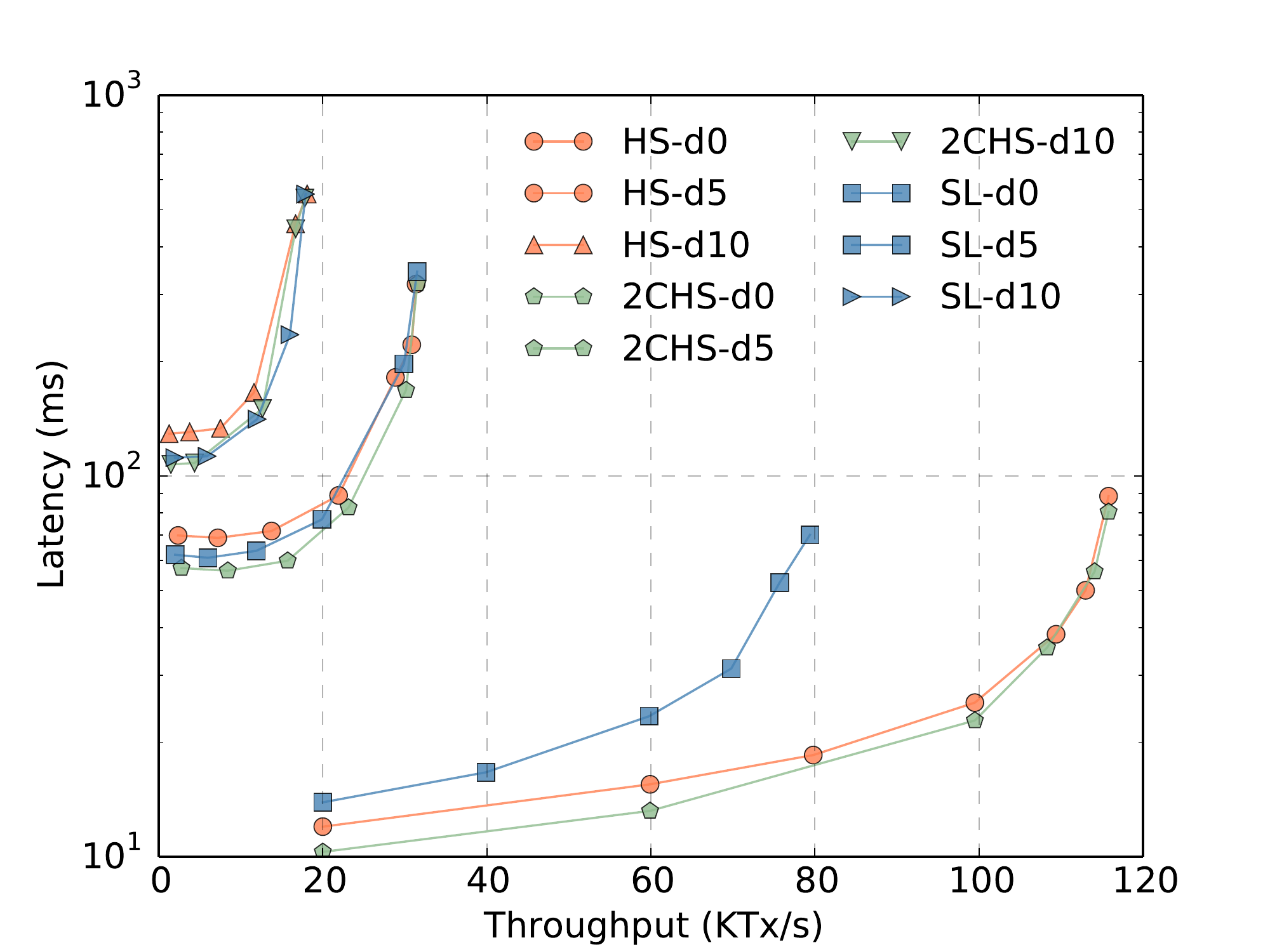}
         \caption{Throughput vs. latency with additional network delays of 0ms, 5ms ($\pm$ 1ms) and 10ms ($\pm$ 2ms).}
         \label{fig:delays}
\end{figure}

\textbf{Network delays.}
We introduce additional network delays between replicas, which are configurable in Bamboo's network module.
Figure~\ref{fig:delays} depicts results under no added network delays and additional network delays of 5ms $\pm$ 1.0ms and 10ms $\pm$ 2.0ms, denoted by ``d0'', ``d5'', and ``d10'', respectively.
We can observe that all the protocols suffer when network delay is increased.
We already know that without added delays, the two HotStuff protocols have the same throughput while there is a clear gap with Streamlet in terms of latency and throughput.
However, when the network delay is 5ms $\pm$ 1.0ms, this gap starts to decrease and at 10ms $\pm$ 2.0ms, Streamlet has comparable performance to 2CHS.
This is because the long network delay compensates for the adverse effect of message echoing in Streamlet.

\begin{figure}[h]
    \begin{center}
      \includegraphics[width=0.9\linewidth]{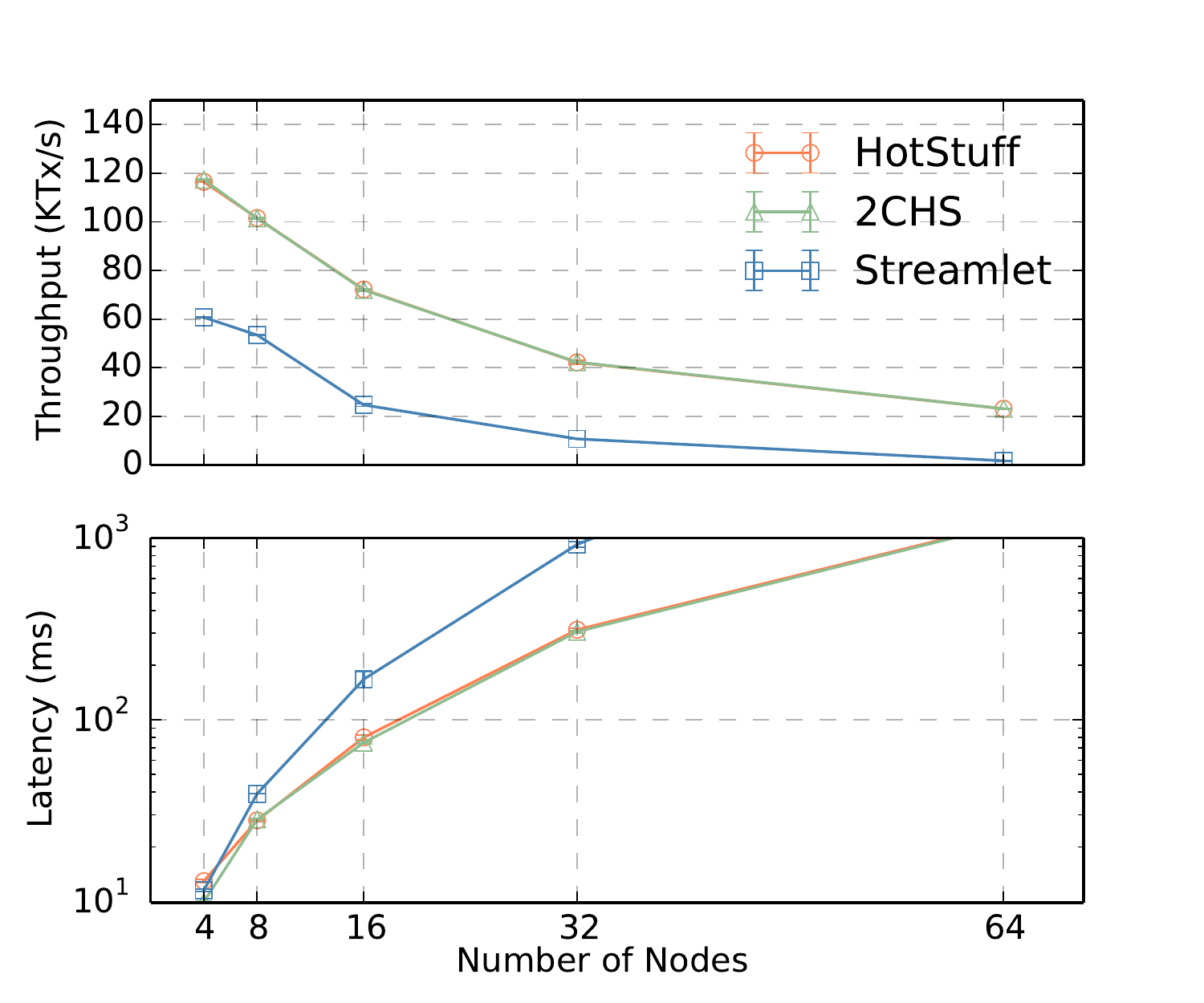}
      \caption{Scalability for 128 byte payload and a block size of 400 txns.}
      \label{fig:scalability}
    \end{center}
\end{figure}

\textbf{Scalability.}
We evaluate the scalability of the three protocols with nodes from the same cluster.
There are no added network delays.
We derive each data point by averaging the results of three repeated runs for 10,000 views under the same setting. We show error bars to present the standard deviation.
The results are plotted in Figure~\ref{fig:scalability}.
Streamlet has the worst scalability and its results obtained at the network size above 64 are meaningless due to high message complexity.
HotStuff and two-chain HotStuff have comparable scalability and the latency difference decreases as the network size increases.
Note that one reason for the sharp increase of latency is the message complexity of the protocols, while another reason is the increase of the waiting time for leader election in which each replica has a local memory pool for storing transactions. 

\subsection{Byzantine Strategies}
To evaluate the ability to tolerate Byzantine attacks discussed in Section~\ref{sec:byzantine}, we conduct two experiments, one for each Byzantine strategy.
Besides throughput and latency, we also measure two micro metrics chain growth rate (CGR) and block intervals (BI) described in Section~\ref{sec:metrics}.
This adds another dimension for considering the results while avoiding the influence of timeout settings.
The transactions contained in the forked blocks are collected back and inserted at the front of the queue of the memory pool.
In the two experiments, we run 32 nodes in total with an increasing number of Byzantine nodes.
The rest of the settings are the same as the scalability test.

\begin{figure}[h]
    \begin{center}
      \includegraphics[width=\linewidth]{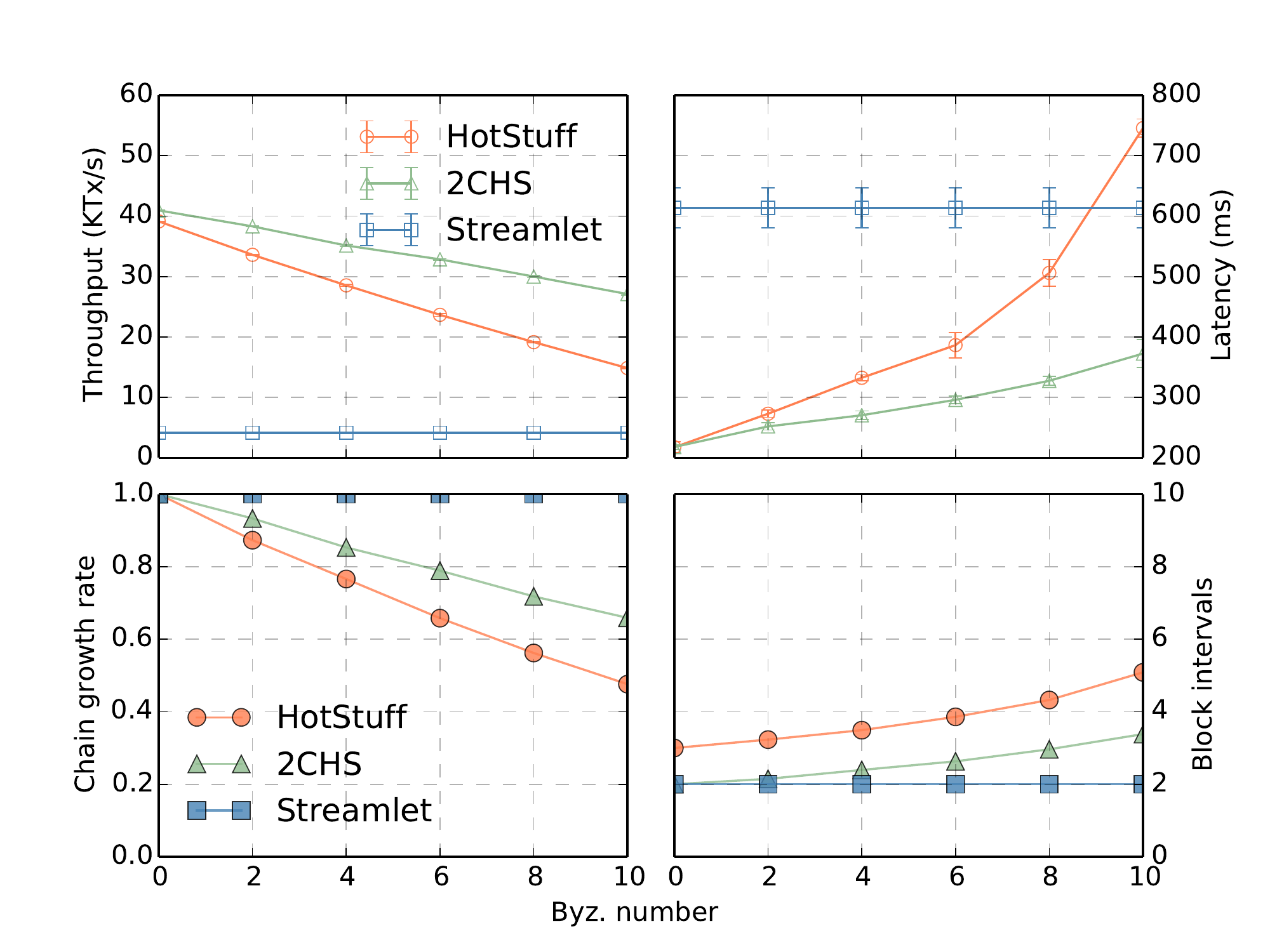}
      \caption{Protocols under forking attack with fixed 32 nodes in total and increasing Byzantine nodes.}
      \label{fig:forking-attack-data}
    \end{center}
\end{figure}

\textbf{Forking attack.}
As described in~\ref{sec:byzantine}, attackers aim to overwrite as many uncommitted blocks as possible.
They propose blocks that could cause forks without violating the voting rule.
Figure~\ref{fig:forking-attack-data} depicts the first setting with Byzantine nodes performing a forking attack.
The line for Streamlet is flat across all the metrics, indicating that it is immune to the forking attack.
This is due to the broadcasting of votes and the longest-chain rule: honest (non-Byzantine) replicas only vote for the longest notarized chain they have ever seen.
Therefore, a fork cannot happen in Streamlet as long as the network is synchronous.
Two-chain HotStuff outperforms HotStuff in all the metrics.
This is because the two-chain commit rule is naturally more resilient against forking in the sense that the attacker can overwrite only one block in 2CHS but can overwrite two blocks in HotStuff.
This also explains why HotStuff and two-chain HotStuff have the same pattern in throughput and CGR.
In terms of block intervals, 2CHS and HotStuff have the same pattern but start from 2 and 3 due to their two-chain and three-chain commit rules, respectively.
However, there is a rapid growth in HotStuff's latency.
This is because the transactions in the forked blocks will be collected back into the memory pool, which increases latency.

\begin{figure}[h]
    \begin{center}
      \includegraphics[width=\linewidth]{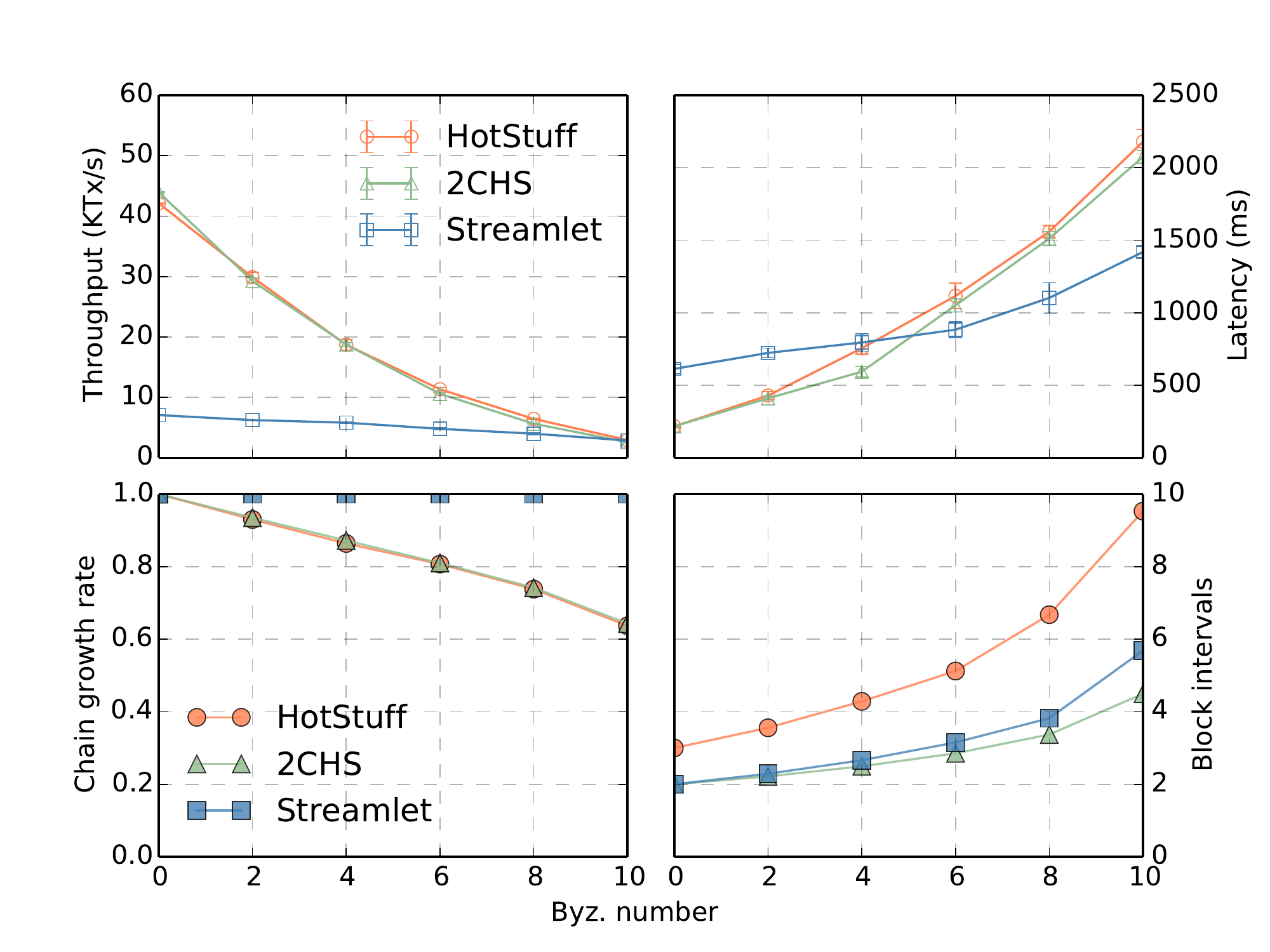}
      \caption{Protocols under silence attack with fixed 32 nodes in total and increasing Byzantine nodes.}
      \label{fig:silence-attack-data}
    \end{center}
\end{figure}

\textbf{Silence attack.}
In the silence attack attackers simply remain silent during their leadership, aiming to break the commit rule.
When the leader is performing a silence attack, the other replicas wait until timeout and try to enter the next view.
In this experiment, we set the timeout to be 50ms to ensure that no additional timeouts are triggered other than the silence attack.
Figure~\ref{fig:silence-attack-data} shows the results of this experiment.
Both HotStuff and 2CHS have the same pattern in CGR and throughput because the silence attack overwrites the last block due to the loss of the QC.
The CGR of Streamlet remains 1 at any run because of the broadcasting of votes and its longest-chain rule so that no fork could happen in a synchronous network.
All protocols have a drop in throughput because of the increasing number of silent proposers.
One note is that although HotStuff and 2CHS show consistently better throughput than Streamlet, Streamlet exhibits graceful degradation due to its immunity to forking.
In terms of block intervals, all protocols show higher BI and more rapid growth than when they are under forking attack in Figure~\ref{fig:forking-attack-data} because for an attacker it is easier to break the commit rule than it is to cause a fork.
Both Streamlet and HotStuff have higher BI than 2CHS because their commit rules are more strict.
In the throughput sub-figure on the top right, although Streamlet shows worse latency than HotStuff and 2CHS when the Byz. number is less than 4; but, it outperforms the other two protocols when the number is above 4.
This is partly due to its immunity to forking.
HotStuff and 2CHS have comparable latency because they have similar CGR.

\subsection{Responsiveness}
To show the advantage of responsiveness, we conduct experiments using fault injection.
In this experiment, we run four nodes under the same setting with the scalability test in Section~\ref{sec:happy-path} except for specific timeout settings for each view.
In the first setting, each protocol's timeout is set to 10ms, denoted by ``t10''.
We let both 2CHS and Streamlet make proposals as soon as $2f+1$ messages are received after view change, which is the same as HotStuff.
In the second setting, the timeouts are changed to 100ms, denoted by ``t100'', and we let all the protocols wait for the timeout if a view change occurs.
We keep the request rate fairly high in both settings.
During the two experiments under two different settings, we manually inject network fluctuation for 10 seconds during which the network delays between nodes fluctuate between 10 and 100 milliseconds.
Additionally, we let one of the nodes perform a silence attack (crash) afterward.

\begin{figure}[h]
    \begin{center}
      \includegraphics[width=0.9\linewidth]{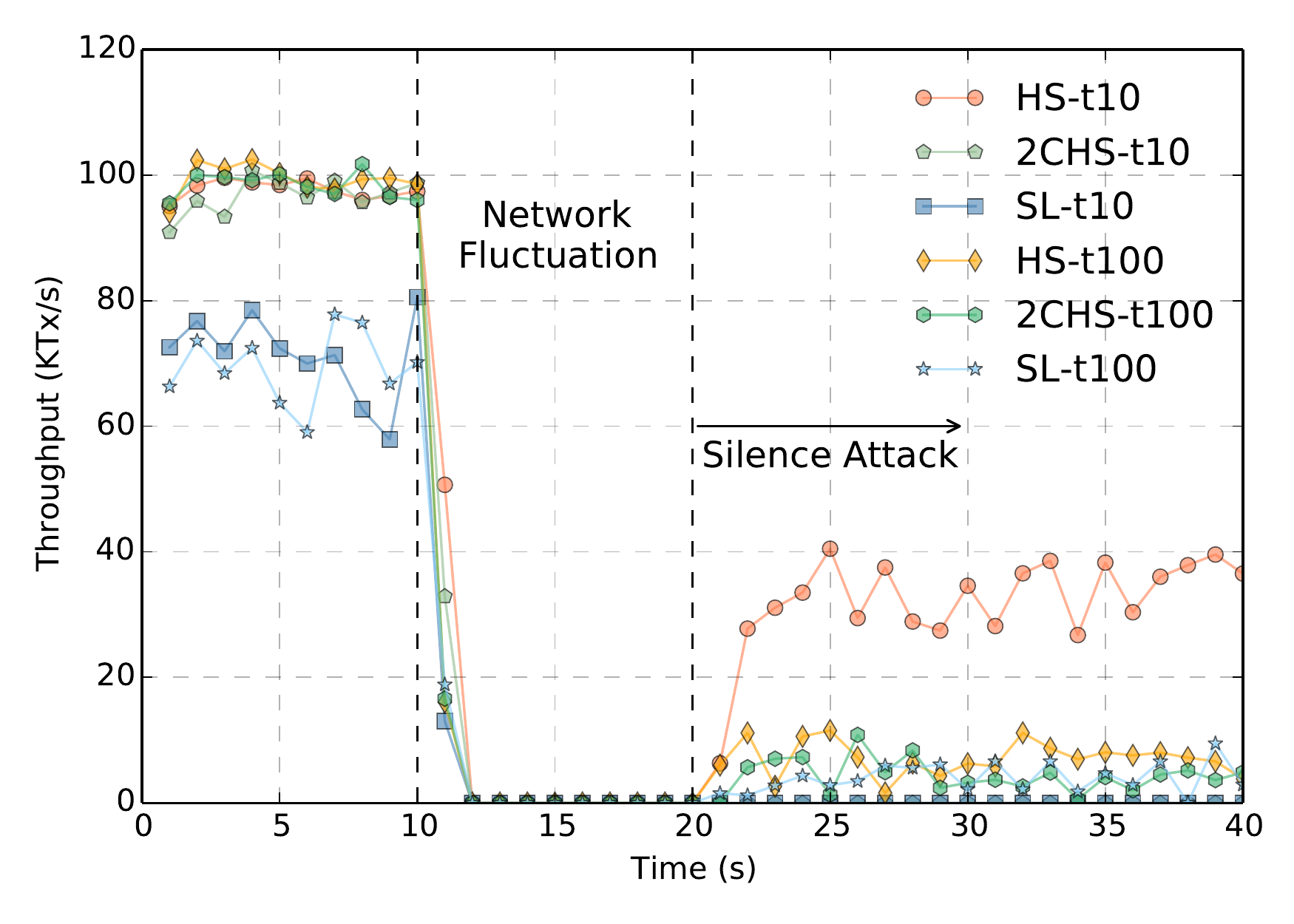}
      \caption{Responsiveness test with timeout settings of protocols of 10ms and 100ms.}
      \label{fig:responsiveness}
    \end{center}
\end{figure}

Figure~\ref{fig:responsiveness} depicts the results from the two settings in one plot.
In the first setting (t10), we observe an abrupt stop of all three protocols during the 10 seconds of network fluctuation.
The HotStuff protocol proceeds instantly after the network fluctuation stops and presents a graceful wave of throughout afterward due to the silence attack.
However, the other protocols never make any progress.
This is because 2 non-Byzantine replicas are locked on a block while the other non-Byzantine replica is locked on a different block during the fluctuation.
When the silence attack is launched, both protocols will not make any progress.

In the second setting (t100), the timeout of each protocol is set to the maximum network delay during the network fluctuation, therefore, the timeout is not triggered until the crashed node is chosen to be the leader.
While all the protocols retain liveness, the throughput is much lower in this setting due to the long timeout.

\subsection{Discussion}
Our experimental results show that the view is also valid in chained-BFT protocols: \textit{one-size-fits-all protocols may be tough, if not impossible, to build}~\cite{BFTsim}.
Different design choices have different trade-offs within a given set of deployment conditions.
In a LAN setup, since HotStuff is one round of voting time slower than 2CHS with comparable throughput, 2CHS might be the better choice.
But this latency advantage is reduced if the transaction has a large payload (e.g., a smart contract deployment).
Optimistic responsiveness makes HotStuff a better candidate for a WAN environment in which the networking conditions are unpredictable.
Whereas the 2CHS protocol, which is not responsive, is subject to the timeout attack~\cite{prime} in a WAN setting.
As for attacks that cause forks, the longest chain rule of Streamlet is more resilient at the cost of high message complexity.
Adapting this rule into HotStuff is an open question and is a good direction for future work in this space.
An alternative way of addressing forking issues is a meticulous incentive design.

\section{Related Work}
Paxi~\cite{Paxi} studies the family of Paxos protocols in a two-pronged systematic approach similar to ours.
They present a comprehensive evaluation of their performance both in local area networks (LANs) and wide area networks (WANs).
BFTSim proposes a simulation environment for BFT protocols that combines a declarative networking system with a robust network simulator~\cite{BFTsim}.
However, although it is handy to have a simulator that enables testing a system in situations that are hard and expensive to reach, we nevertheless need real testing on top of simulation, and BFTSim lacks the ability to produce empirical results.

BFT-Bench~\cite{BFTBench} is the first benchmark framework that evaluates different BFT protocols in various workloads and fault scenarios.
It provides systematic means to inject different networking faults and malicious intent into the system, and compares several important BFT protocols.
However, although the proposed framework allows a modular approach towards load injection and performance evaluation, it does not provide enough granularity and modularity for designing the BFT protocol itself.
This is partly because the protocols that it compares differ in ways that make it hard to capture them in a single abstract model.
In this paper, we leverage the similarities between cBFT protocols to provide a set of common abstractions that make Bamboo possible.

BlockBench~\cite{blockbench} is the first evaluation framework for analyzing existing private blockchains.
It dissects a blockchain system into different layers: consensus, data model, execution, and application.
BlockBench provides both macro and micro workloads for testing different layers.
By contrast, Bamboo only focuses on the consensus layer and provides identical implementations of the other layers.
ResilientDB~\cite{resilientdb} offers insights into optimizing the performance of a permissioned blockchain system via a well-crafted architecture, which can be applied in Bamboo to provide universal optimizations for cBFT protocols.

At the final stage of writing this paper, we noticed a concurrent work~\cite{sft} that develops an abstraction of cBFT protocols similar to ours.
The main contribution of \cite{sft} is a new chained-BFT protocol called SFT to strengthen the ability of BFT SMR to tolerate Byzantine faults. In our future work we plan to express and evaluate this protocol with Bamboo.

\section{Conclusion and Future Work}
We present a prototyping and evaluation framework called Bamboo, for comparing chained-BFT protocols.
We also introduce a mathematical model to validate our experimental results, which also helps to distill the performance of cBFT protocols.
We build multiple cBFT protocols using Bamboo and present a comprehensive evaluation of three representatives under various scenarios, including two Byzantine attacks.
Our results echo the belief that there is no single best BFT protocol for every situation.
Bamboo allows developers to analyze cBFT protocols in light of their fundamental trade-offs, and build systems that are best suited to their context. 
In our future work, we will use Bamboo to further explore cBFT protocol designs, focusing specifically on protocols with relaxed consistency guarantees.

\bibliographystyle{unsrt}
\bibliography{paper}

\end{document}